\documentclass{elsart}
\usepackage{epsfig}
\newcommand{\gvec}[1]{\hbox{\boldmath$#1$\unboldmath}}  

\newcommand{\ket}[1]{|#1 \rangle}
\newcommand{\bra}[1]{\langle #1|}

\begin{document}

\vskip 1 true cm

\begin{flushleft}
LYCEN 99128 \\
CERN-TH/99-362
\end{flushleft}

\begin{frontmatter}
\title{Pion scalar density and chiral symmetry restoration 
 at finite temperature and density}
 
\author{G.~Chanfray$^a$, D.~Davesne$^a$, J.~Delorme$^a$, 
M.~Ericson$^{a,b}$ and J.~Marteau$^a$}                              
\address{$^a$Institut de Physique Nucl\'eaire de Lyon, IN2P3-CNRS et
Universit\'e Cl. Bernard, \\
43, Bvd. du 11 novembre 1918, F-69622 Villeurbanne Cedex, France}
\address{$^b$ CERN,  Theory Division, CH-1211 Geneva 23, Switzerland}

\begin{abstract}
This paper is devoted to the evaluation of the pionic scalar density at finite
temperature and baryonic density. We express the latter effect in terms of the
nuclear response evaluated in the random phase approximation. We discuss the
density and temperature evolution of the pionic density which governs the quark
condensate evolution. Numerical evaluations are performed.

\end{abstract}
\end{frontmatter}

 
\section*{Introduction}

Pions play a crucial role in chiral symmetry restoration, due to their Goldstone
boson character. The amount of restoration is measured by the modification
of the order parameter, i.e. the quark condensate density, with respect to the
vacuum value. In this context the 
evaluation of the expectation value of the squared pion field, linked to the
scalar density of pions $\rho ^\pi_S$ by $\rho ^\pi _S =m_\pi  \langle \Phi^2
\rangle$, is of a great interest.
For a single nucleon this quantity governs the total amount of chiral symmetry
restoration of pionic origin, according to~\cite{CE,CEW}:
\begin{equation} \label{eq1}
\frac{m_\pi^2}{2} \int d^3\vec x \bra{N} \Phi^2(x)
\ket{N} = 2 m_q \int d^3\vec x \bra{N} \Delta ^\pi \overline q q(x) \ket{N} 
= \Sigma^\pi_N = {m_\pi\over 2} N_\pi,
\end{equation}
where  $N_\pi$ is the scalar number of pions in the nucleon cloud, 
$\Sigma^\pi_N$ is the part of the nucleon $\Sigma$ commutator of pionic origin
and $\bra{N} \Delta ^\pi \overline q q(x) \ket{N}$ represents the
corresponding modification of the quark condensate with respect to the
vacuum value.
Similarly, in a uniform nuclear medium of density $\rho$, or in a 
heat bath, the evolution of the quark condensate originating from the pions
is linked, to one-pion loop order, to the average value  $\langle \Phi^2 \rangle$ by~:
\begin{equation}\label{eq2}
\frac{\Delta ^\pi \langle \bar{q} q (\rho ,T) \rangle}{\langle \bar{q} q (0,0)
\rangle} = - \frac{\langle \Phi^2 \rangle}{2f_\pi^2}\,. \label{qqbar}
\end{equation}

The same quantity $\langle \Phi^2 \rangle$ governs also the quenching factor,
$1-\langle \Phi^2 \rangle/3f_\pi ^2$, of coupling constants such as the
nucleonic axial coupling constant $g_A$ or the pion decay one $f_\pi$,
originating from pion loops, which is the counterpart of the mixing of axial
and vector currents \cite{IOF,CDE1}. This type of
quenching has to be seen as a manifestation of chiral symmetry restoration and
should also apply to the case of $\rho$ meson excitation by virtual photons as
enters in relativistic heavy ion collisions. It is therefore interesting to
evaluate the quantity $\langle \Phi^2\rangle$ in the conditions of such
experiments. Now, the fireball which is the source of the dileptons 
contains, besides thermal pions, a significant residual baryonic background
We have therefore to understand how the pion density evolves at finite values
of both temperature and baryonic chemical potential. The first order
approximation for the quantity $\langle \Phi^2 \rangle $ adds the values for a
pure heat bath and for a cold baryonic medium, writing with obvious notations~:
\begin{equation}
\langle \Phi^2 \rangle(\rho , T) = \langle \Phi^2\rangle_T (\rho=0)
+ \langle \Phi^2\rangle_\rho (T=0)  \,.
\label{addit}
\end{equation}
However this approximation is likely to be crude since the 
temperature has an effect on the pion density of the 
nuclear medium and on the other hand the presence of the 
baryonic background modifies the number of pions thermally 
excited.  As an illustration of the second point, the pion density 
in the baryonic vacuum is fixed by the Bose-Einstein factor which is
$\left ( e^{\omega _k/T} -1 \right ) ^{-1}$ for pions of momentum $\vec k$, 
with $\omega _k= \sqrt{\vec k ^2 + m_\pi ^2}$. In 
the nuclear medium the pion becomes a quasi-particle with a 
broad width. It can decay for instance into a particle-hole pair 
which has a smaller energy than the free pion. Its excitation is 
then favored by the thermal factor. There is therefore a mutual 
influence between temperature and density that we will 
investigate.

The article is organized as follows. In the first section we evaluate the pion
density of a nuclear medium at zero temperature. We relate this quantity to the
nuclear response to a pion-like excitation. We evaluate this response in the
RPA scheme, taking also into account the two particle-two hole ($2p$-$2h$)
 excitations. Within this
framework, we study the deviation with respect to the independent nucleon 
approximation. In the second section we introduce the effect of the temperature
through the modification of the nuclear responses. In the third section we
incorporate the influence of the finite baryonic chemical potential in the heat
bath case.

\section{Pion scalar density in the cold nuclear medium}

\begin{figure}[h]
\begin{center}
\includegraphics[width=8 true cm,height=2 cm]{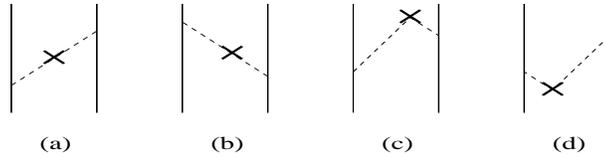}
\end{center}
\caption{\label{fig1} \textit{Time-ordered graphs for pion exchange.}}
\end{figure}

This quantity was discussed in relation to the quark condensate 
modification by Chanfray and Ericson \cite{CE}. They discussed
its deviations with respect to free nucleons, introducing the nuclear 
response to pion-like excitation, treated in the static case $M_N \rightarrow
\infty$. The extension to the non-static situation  can be performed  
through the time-ordered graphs of Fig.~\ref{fig1} where the 
cross represents the point at which the pions are created or 
annihilated. The pion momentum is denoted $q$ and $\omega$ is the 
excitation energy of the nuclear system in the intermediate 
state. The sum of the four graphs leads to the following 
expression which has also been derived with a different method in
Ref.~\cite{CD}~:
\begin{eqnarray} Eq.
\label{eq6}
\langle \Phi^2 \rangle = \frac{\rho}{A} \, 3 \, \frac{g_{\pi NN}^2}{4M_N^2} \, 
\int \frac{d^3q}{(2\pi)^3 \, 2 \omega_q^2} \int_0^\infty d\omega \left[ 
\frac{1}{(\omega+\omega_q)^2} + \frac{1}{\omega_q}\frac{1}{(\omega+\omega_q)} 
\right] \, v^2(Q^2) \,R_L(\omega,\gvec{q}) 
\end{eqnarray}
where $Q^2 =\omega^2-\gvec{q}^2$, $v(Q^2)$ is the form factor of the pion
vertex for which we use a monopole form~:
\begin{equation} \label{formfactor}
v(Q^2) = \frac{\Lambda^2 - m_\pi^2}{\Lambda^2 - Q^2}.
\end{equation}
Finally, $R_L$ represents the spin-isospin longitudinal response function~:
\begin{equation} 
R_L(\omega,\gvec{q})=
\sum_n\,\vert\langle n|\sum_{i=1}^A\,\sigma_i.\gvec{q}\, \tau_i^a
\,e^{i \gvec{q}.\gvec{r}_i}|0\rangle\vert ^2\,\delta(\omega-E_n).
\end{equation}

\noindent In the actual calculation, we add in the response $R_L$ the
excitation of the Delta resonance with the standard replacement~:
\begin{equation}
\gvec{\sigma}_i.\gvec{q} \,\, \tau^a_i \longrightarrow
\frac{g_{\scriptscriptstyle \pi N\Delta}}{g_{\scriptscriptstyle \pi NN}} \,
\gvec{S}^\dagger_i.\gvec{q} \,\, (T^a_i)^\dagger 
\end{equation}
where $\gvec{S}^\dagger$ ($\gvec{T}^\dagger$) is the spin (isospin) transition
operator
connecting the spin (isospin) $\frac{1}{2}$ and $\frac{3}{2}$ states 
\cite{EW}.
Note that for a matter of convenience we have incorporated in this operator
the ratio of the $\pi N\Delta$ and $\pi NN$ coupling constants. 
 We have also
assumed the same form factors, $v(Q^2)$, at the $\pi NN$ vertex and at the
$\pi N\Delta$ vertices.
 Finally we recall the link between the response function
and the polarization propagator $\Pi(\omega,\gvec{q},\gvec{q'})$ which we will
use in the following~:
\begin{equation} \label{resp}
R(\omega,\gvec{q}) = -\frac{V}{\pi} \, \mathrm{Im} \Pi(\omega,\gvec{q},
\gvec{q}) 
\end{equation}
We will first discuss the result for the free nucleon.

\subsection{Free nucleon.}

In the nucleon case, where the response $R_L$ reduces to a simple expression,
Eq.~(\ref{eq6}) provides for the (scalar) pion number $N_\pi$ in the nucleon
cloud~:
\begin{eqnarray} 
N_\pi & = & m_\pi\int d^3 \vec x \bra{N} \Phi^2 ({\bf x}\ket{N} \nonumber \\
      & = & m_\pi \, 3 \, \frac{g_{\pi NN}^2}{4M^2} \, \int
            \frac{d^3q}{(2\pi)^3} \gvec{q}^2 \left\{\frac{1}{\omega_q} \left[
	    \frac{1}{2\omega_q} \frac{1}{(\varepsilon_q+\omega_q)^2} +
	    \frac{1}{2\omega_q^2}\frac{1}{(\varepsilon_q+\omega_q)} \right]
	    \, v^2(Q^2) \right. \nonumber \\
      &   & \left. + \frac{4}{9} \, (\frac{g_{\scriptscriptstyle \pi N\Delta}}
            {g_{\scriptscriptstyle \pi NN}})^2 \int_0^\infty \frac{d\omega}{\omega_q} 
            \left[ \frac{1}{2\omega_q} \frac{1}{(\omega+\omega_q)^2} +
            \frac{1}{2\omega_q^2}\frac{1}{(\omega+\omega_q)} \right] \,
	    v^2(Q^2)\left(- {1\over \pi}\mathrm{Im} 
	    \frac{1}{\omega-\omega_\Delta 
	   + i\frac{\Gamma_\Delta}{2}}\right)\right\}\,.
\end{eqnarray}
In the above equation we have used the following notations~: $\varepsilon_q =
\gvec{q}^2/2M_N$ and $\omega_\Delta = M_\Delta - M_N + \gvec{q}^2/2M_\Delta$.
The energy dependence of the Delta width $\Gamma_\Delta$ is taken from the
analysis of the pion-nucleon scattering \cite{EW}.

Our numerical inputs for the evaluation of $N_\pi$ are defined as follows.
Without form-factor the integrals
diverge linearly. The resulting value is
then quite sensitive to the cut-off function. We stress, however, that the
specific case of the single nucleon is not the purpose of this paper. It is
for us an element of
comparison to introduce the nuclear effects. Since we want to restrict the
calculation to a region where nuclear effects are reasonably under control, we
have limited the integrals to~: $q = 1$ GeV and $\omega = 1$ GeV.
The parameters are chosen so as to obtain a value $\Sigma^\pi_N = 
\frac{1}{2} m_\pi N_\pi = 30$ MeV which is well in the accepted range
\cite{BMG,JAM,CDE2}. This is achieved with $\Lambda = 1$\,GeV 
and $(g_{\scriptscriptstyle \pi N\Delta}/g_{\scriptscriptstyle \pi NN})^2 = 3.8$.

\subsection{Infinite nuclear matter.}

\begin{figure}[h]
\begin{center}
\includegraphics[width=16.5 true cm,height=4cm]{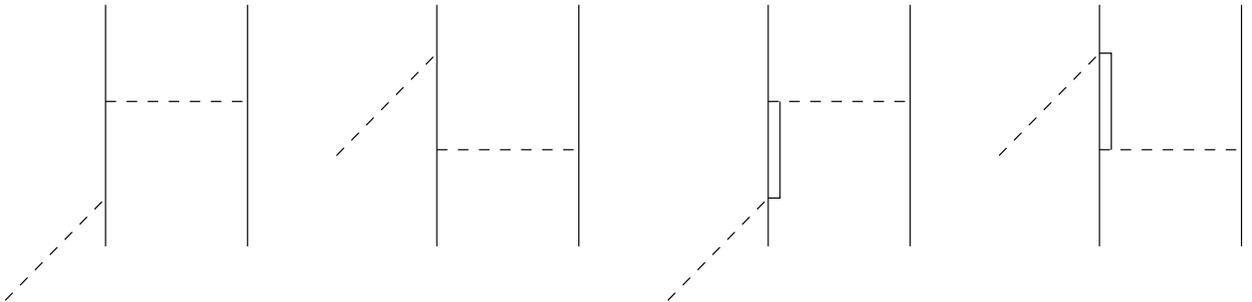}
\end{center}
\caption{\label{fig2} \textit{Feynman diagrams for the $2p$-$2h$ processes. The
double line represents the Delta resonance.}}
\end{figure}

We now turn to the case of infinite nuclear matter.
In order to evaluate the response functions, we use the method of
Delorme and Guichon \cite{DG} who calculated the zero order response in the
local density approximation and then solved exactly the RPA equations. 
Their zero order response function also includes the $2p$-$2h$ excitations.
The corresponding Feynman diagrams of such processes are displayed in
Fig.~\ref{fig2}. 

They are calculated in two steps. First we single out the contributions 
which reduce to a medium modification of the $\Delta$
self-energy, for which the parametrization of Ref.~\cite{OST} is used.
This parametrization includes some $3p$-$3h$ excitation states as well.
For the rest we use the results of Shimizu-Faessler~\cite{SF}
who evaluated the two nucleons p-wave pion absorption at threshold
$(\omega = m_\pi)$, from which the $\Delta$
self-energy part, already taken into account, is separated out. As for each the
remaining contributions, an energy extrapolation suggested by the
many-body diagrammatic interpretation is performed.

\begin{figure}[h]
\begin{center}
\includegraphics[width=9 true cm]{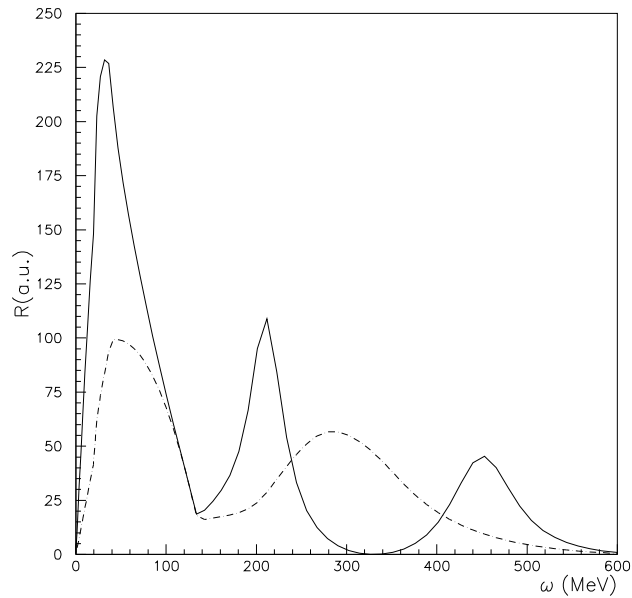}
\end{center}
\caption{\label{fig3} \textit{Response function per nucleon at normal nuclear
density as a function of the energy for a fixed momentum $q$ = 300 MeV.
The dot-dashed and continuous lines represent respectively the zero order and RPA
responses.}}
\end{figure}

We solve the RPA equations in the ring approximation~: $\Pi = \Pi^0 + 
\Pi^0 \, {\mathcal {V}} \, \Pi$. Here ${\mathcal {V}}$ is the particle-hole
($p$-$h$) interaction with the standard formulation~:
 ${\mathcal {V}} = V_\pi + V_{g'}$
where the second piece is the short-range Landau-Migdal part.
More explicitely with our definition of the response, ${\mathcal {V}}$ reads~:
\begin{equation}      
{\mathcal {V}} = {v^2(Q^2)\over \gvec{q}^2}\left ({\gvec{q^2}\over 
Q^2-m^2_\pi}+g'\right)\,. 
\end{equation}
The corresponding
Landau-Migdal parameters $g'$ are different in the various channels~:
${g'}_{NN}$ for the $NN$ sector, ${g'}_{\Delta\Delta}$ for the $\Delta$ one,
${g'}_{N\Delta}$ for the mixing of $N N$ and $\Delta N$ excitations. We adopt
the following values:
\begin{equation}      
{g'}_{NN} = 0.7,\,\,\, {g'}_{N\Delta} = {g'}_{\Delta\Delta} = 0.5\,. 
\end{equation}
The results are illustrated on Fig.~\ref{fig3} which shows the energy 
dependence of the zero order and RPA responses, for a
fixed value of the momentum $q = 300 $\,MeV. The bare response presents a
low-energy peak corresponding to the $NN^{-1}$ excitation and a high-energy
one ($\Delta N ^{-1}$ excitations).
The figure displays the RPA enhancement of the low energy peak, introduced by
Alberico et {\it al.} \cite{AEM}, which arises from the attractive
nature of the $p$-$h$ interaction. It also displays the collective behaviour of
the $\Delta$ excitation with a splitting into two branches.

Once the energy and momentum integrations are performed, we find, as in
 Ref.~\cite{CE},
 a moderate increase as compared to the free nucleon value,
 with the values per nucleon at normal nuclear density~:
\begin{equation}
\tilde\Sigma^\pi_N \equiv\Sigma^\pi_A/A = 38.5 \,\,\, \mathrm{MeV} \,\,\,
 \mathrm{and} \,\,\, N^\pi_{A}/A = 0.55\, 
\end{equation}
versus $30$\,MeV and 0.40 respectively for the free nucleon. 
 The quantity $\tilde\Sigma^\pi_N$ is the (medium modified) effective 
sigma commutator.      

\begin{figure}[h]
\begin{center}
\includegraphics[width=16.5 true cm,height=4cm]{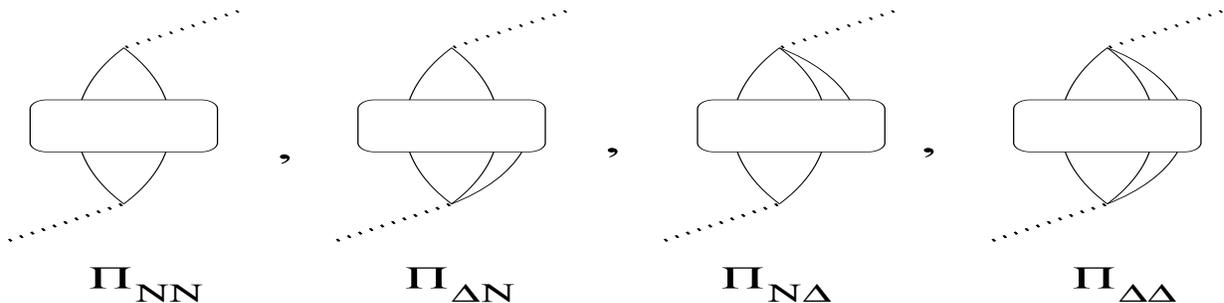}
\end{center}
\caption{\label{fig4} \textit{Symbolic representation of the $NN$, $\Delta N$, 
$N\Delta$and $\Delta\Delta$ response functions. The Delta resonance is 
represented by the double line.}}
\end{figure}

\begin{figure}[h]
\begin{center}
\includegraphics[width=12 true cm,height = 10 cm]{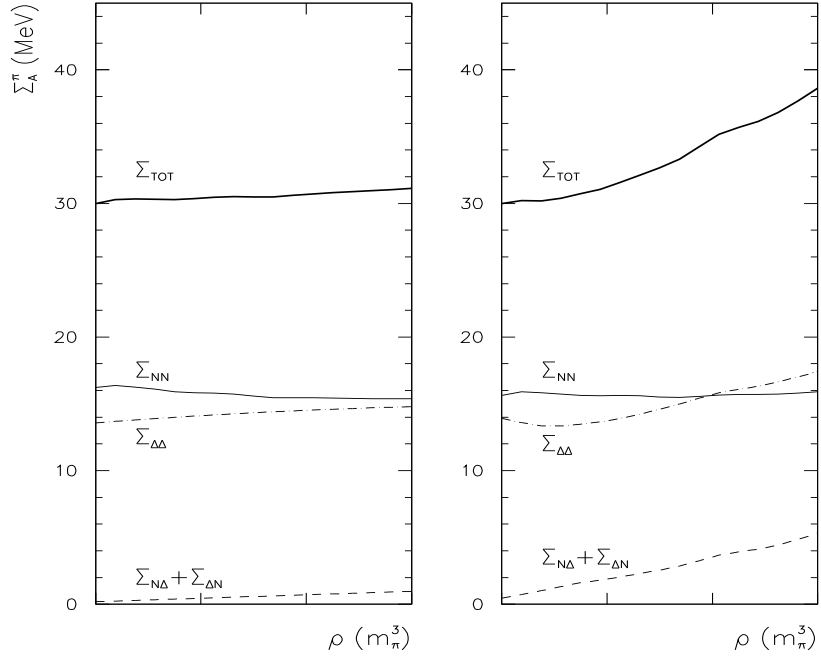}
\end{center}
\caption{\label{fig5} \textit{Density evolution of the different components of
the effective sigma commutator: left (right) figure is without (with) RPA.}}
\end{figure}

More precisely, we decompose the response function $\Pi$ (accordingly
$\Sigma ^\pi$) into four types, depending on the kind of states which are
excited at each external vertices~: $NN$, $N\Delta$, $\Delta N$ and
$\Delta\Delta$ (Fig.~\ref{fig4}). 
The density evolution of the different components of the sigma commutator
are represented on Fig.~\ref{fig5} both without and with RPA. Notice that,
in absence of the RPA, there
is already a contribution of the $N\Delta$ channel at finite density due to the
$2p$-$2h$ excitations. The overall RPA increase of the sigma commutator mainly
comes from that of the $\Sigma_{N\Delta} + \Sigma_{\Delta N}$ parts. These last
quantities embody the mixing of the $\Delta N^{-1}$ configurations into the
$NN^{-1}$ ones. This is well known to be responsible for the enhancement
of the low energy response ({\it i.e.} the $NN^{-1}$ excitations)~\cite{ME}.
\section{Inclusion of temperature} 

\subsection{Influence on the virtual pion cloud}

We now introduce the temperature, {\it via} the Matsubara formalism~: all the
integrals
over energy are replaced by an infinite sum over Matsubara frequencies.
For the $NN^{-1}$ sector, the generalized Lindhart function
(the imaginary part of which is proportional to the response function) 
 at finite temperature is discussed in textbooks (see e.g. Ref.~
\cite{FW}). The result is the replacement of the Heaviside
functions that characterize the occupation of
fermion states at zero temperature by the Fermi-Dirac distributions. 
We have generalized this procedure to particle with width and applied it to
$\Delta$-$N^{-1}$ , $N$-$\Delta^{-1}$ and $\Delta$-$\Delta ^{-1}$ rings. 
The generalized Lindhart
function $L_{ab^{-1}}$ for a process involving a particle of type (a) and a hole of
type (b) is found to be~:
\begin{equation}
L_{ab^{-1}}(\omega, \vec q) = - \frac{N_{S,I}}{4\pi^2}\int k^2 dk d(\cos
\theta) \frac{f(\omega _k ^b)(1-f(\omega _{k+q} ^a))}{\omega -\omega _{k+q}
 ^a + \omega _k ^b +\frac{i}{2}\Gamma ^a (\omega _k ^b + \omega) 
 + \frac{i} {2}\Gamma ^b
(\omega _{k+q} ^a- \omega)}
\end{equation}
\begin{equation}
- \frac{f(\omega _k ^b)(1-f(\omega _{k+q} ^a))}
{\omega +\omega _{k+q} ^a
- \omega _k ^b +\frac{i}{2}\Gamma ^a (\omega _k ^b - \omega) + 
\frac{i} {2}\Gamma ^b
(\omega _{k+q} ^a + \omega)}
\end{equation}
where $N_{S,I}$ is a constant arising from  the summation of spin and isospin
and $\Gamma ^{a,b} (\omega)$ represents the width of the particle of type (a,b) for
an energy $\omega$. The occupation number  of  hadron species $a$ is~:
\begin{equation}
f(\omega^a_k)={1\over exp({\omega^a_k-\mu\over T})+1}
\end{equation}
where $\mu$ is the (common) chemical potential for baryons, the value of which
 fixes the
baryonic density $\rho$ at a given temperature. As implicitely stated before, we limit
ourselves to nucleons and deltas {\it i.e.} $\rho=\rho_N+\rho_\Delta$. 

Concerning the $\Delta$ width, things are somewhat more complicated.
In the medium, the pionic decay channel $\Delta \rightarrow \pi \, N$ is partly
suppressed due to Pauli blocking. At the same time, other channels open, the
pion being replaced by $1p$-$1h$, $2p$-$2h$, etc... At normal density and zero
temperature the pionic channel remains dominant according to Ref.~\cite{OST}.
 It represents approximatively 75\% of the total width ($\simeq 90$ MeV to be
compared with 120 MeV at the resonance energy) and the non-pionic decay
 channels only the remaining 25\%. In
view of the difficulties of  a full calculation of  the
temperature effects, we have adopted the simplified following strategy~:
we have kept the parametrization of Ref.~\cite{OST}, derived at
$T = 0$, for the non-pionic decay channel. We have introduced the  temperature effects
{\it via} the Matsubara formalism only for the main pionic part of the width.

\begin{figure}[h]
\begin{center}
\includegraphics[width=10 true cm]{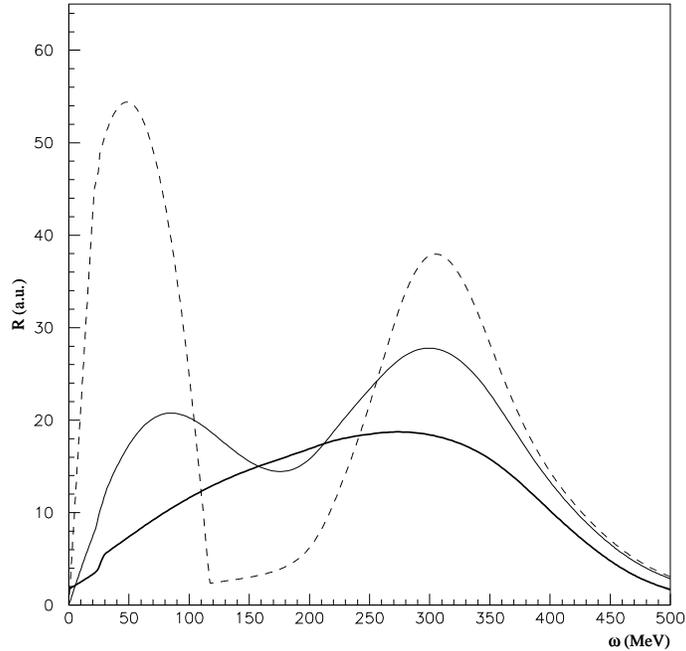}
\end{center}
\caption{\textit{Zero order response at a baryonic density equal 
half nuclear matter density 
($\rho=0.25 m^3_\pi$) as a function of energy for a fixed momentum $q=300$ \, MeV  
for three
temperatures.  Dashed lines~: $T=0$. Thin full line~: $T=50$\, MeV. Thick full line~:
$T=150$ \, MeV.}}
\label{tfinienu}
\end{figure}

\begin{figure}[h]
\begin{center}
\includegraphics[width=10 true cm]{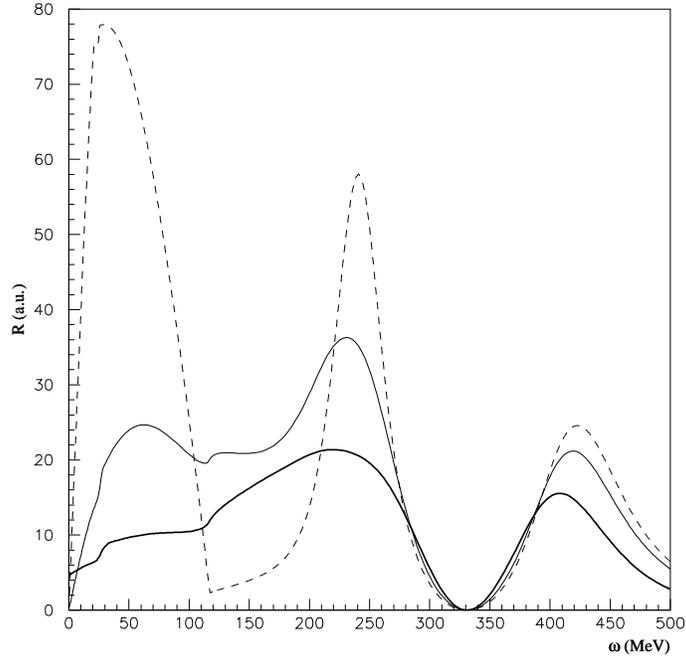}
\end{center}
\caption{\textit{Same as before but for the RPA response.}}
\label{tfinierpa}
\end{figure}

The Figure~\ref{tfinienu} displays the
temperature evolutiom  of the zero order response function. The dashed line, 
which represents the $T=0$
case, exhibits the $N$-$N^{-1}$ and $\Delta$-$N^{-1}$ structures. Increasing 
temperature tends to wash out more and more these peaks. At the same time an overall
suppression effect occurs.
Note that the $\Delta$ branch is less affected by the temperature because of
the higher energies involved. 
In the RPA case (see Fig.~\ref{tfinierpa}) , one observes a similar behavior~:
an important general decrease and the loss of the lower energy structures. 

We now turn to the question of the thermal pions present in the heat bath.

\subsection{Inclusion of thermal pions.}

The  effects previously discussed  concerned only the modification, due to the
temperature, of the virtual pion density present in the nuclear medium. At finite
temperature, thermally excited pionic modes (quasi-pions) are also present 
and they give an additional contribution to the pion scalar density. Here for
a better illustration we give the result for $\langle \Phi^2\rangle/2 f^2_\pi$ 
which according to Eq.(\ref{qqbar}) governs the amount of chiral symmetry
 restoration of pionic origin: 
\begin{eqnarray*}
{\langle \Phi^2\rangle\over 2 f^2_\pi}
       & = & {\rho\over A}\frac{3}{2f_{\pi}^2} {g^2_{\pi NN}\over 4 M_N^2}
              \int{d^3 q\over (2\pi)^3}\, \int_0^\infty d\omega\,
             \left({1\over 2\omega^2_q (\omega+\omega_q)^2}\,+ \,
	     {1\over 2\omega^3_q (\omega+\omega_q)}\right)\,v^2(Q^2)\,R_L(\omega,
	     \gvec{q})  \\ 
       &   & + \frac{3}{2f_{\pi}^2} \int{d^3 q\over (2 \pi)^3}\,\int_0^\infty d\omega\,
             n(\omega)\,\left(-{2\over \pi}\right)
                              \mathrm{Im}\,D(\omega,\gvec{q})   \\ 
       & \equiv & \frac{\rho \tilde \Sigma_B^{\pi}}{f^2_\pi m^2_\pi}+ 
       \frac{<\Phi^2>_T}{2f_{\pi}^2}
\end{eqnarray*}
where $n(\omega)=1/\left(exp(\omega/T) -1\right)$ is the Bose occupation factor
and $D(\omega,\gvec{q})$ the quasi-pion propagator. 
The second identity defines the quantities $\tilde\Sigma_B^{\pi}$
 and $<\Phi^2>_T$, associated respectively with the first and
 second pieces of the r.h.s. of the above 
equation: $\tilde\Sigma_B^{\pi}$ represents an effective, 
temperature dependent, sigma commmutator per baryon, whereas $<\Phi^2>_T$ is
 the scalar density of quasi-pions thermally excited.

\section{Results}

In order to display the condensate evolution, all the forthcoming figures show 
its relative decrease, {\it i.e.} the quantity $ \Phi^2/2f_\pi^2$  according to
Eq.~(\ref{eq2}). We stress
again that the points of interest are the influence of temperature on the 
nuclear pionic cloud contribution, the influence of the baryonic density on the
thermal pions one and finally how large is the deviation from the additive
approximation of Eq.~(\ref{addit}). 
  
\begin{figure}[h]
\begin{center}
\includegraphics[width=8 true cm]{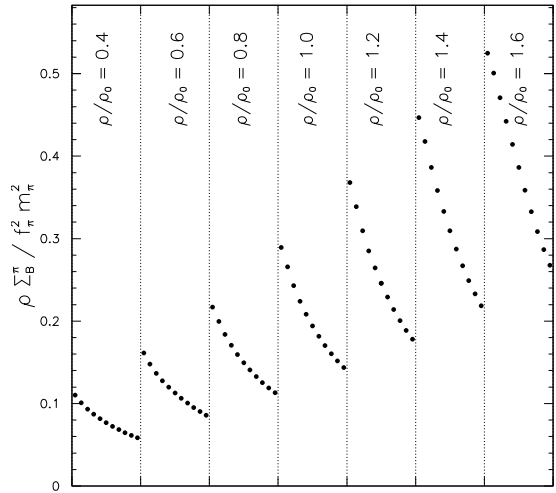}
\end{center}
\caption{\textit {Relative decrease of the condensate coming from the virtual
 pion cloud part as a function of the density and the temperature. Each box 
 corresponds to a fixed density as 
indicated. The density increases by steps of $0.2 \rho_0$ from left to right
 between $0.4$ and $1.6 \rho_0$. In each box the points
correspond to temperature increases by steps of $0.10 m_\pi$ between $0.05$
 and $1.05 m_\pi$. }}
\label{sigrpa}
\end{figure}

We first present on Fig.~\ref{sigrpa} the contribution of the virtual pion cloud alone 
alone (term in $\tilde\Sigma_B^{\pi}$). Each box shows the temperature evolution at
 fixed baryonic density. It illustrates the suppression effect of the
 temperature which originates in the quenching of the nuclear response
 previously mentioned. The increase with baryonic density observed in this
 figure reflects the obvious fact that the pionic density follows the baryon 
 one.  
 
\begin{figure}[h]
\begin{center}
\includegraphics[width=8 true cm]{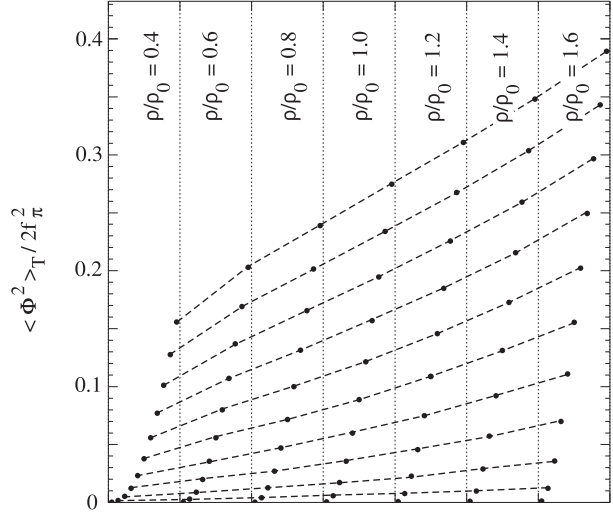}
\end{center}
\caption{\textit{Same as the preceding figure but for the thermal pions.
As a guidance in order to display the influence of the density on the thermal
pions, points of equal temperature are joined by a dashed line.}}
\label{sigphi}
\end{figure}

In Figure \ref{sigphi} we present in the same fashion the condensate decrease
due to thermal pions alone (term in $\langle \Phi^2\rangle_T/2f_{\pi}^2$). 
The iso-temperature curves show the influence of the baryonic density which
pushes down the quasi-pion excitation energy, thus increasing their thermal
excitation. 

\begin{figure}[h]
\begin{center}
\includegraphics[width=8 true cm]{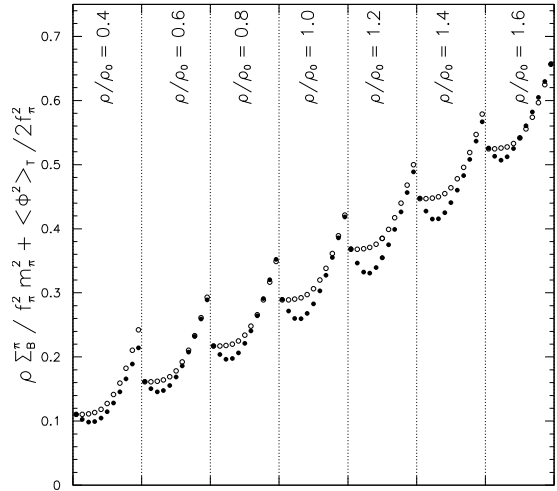}
\end{center}
\caption{\textit{Sum of both contributions of the virtual pion cloud and the
thermal pions (black circles). For comparison the open circles represent the
additive approximation of \protect Eq.~(\ref{addit})}}
\label{sigtot1}
\end{figure}

Finally in the last figure (Fig.~\ref{sigtot1}) we present the sum of both
contributions. The competition between the variations of both terms with respect
 to the temperature is the source of the observed parabolic type shape. For
 comparison, we have plotted in open circles the approximation of
 Eq.~(\ref{addit}) where the effect of the thermal pions at zero density is
 simply added to that of the pion cloud at zero temperature. 
 This approximation does not display the hollow shape of the exact calculation:
 it overestimates the latter by at most 15\% in the medium part of
 the temperature range we have considered, the deviation becoming quite small 
 beyond $T \approx$ 90-100 MeV. In this region the decrease of the pionic
 cloud contribution with temperature (Fig.~\ref{sigrpa}) practically compensates
 the enhancement of the thermal excitations by density effects. The most
 important conclusion which can be drawn from Fig.~\ref{sigtot1} is that, due to
 the nuclear pions, the pion scalar density is much larger than in the absence
 of nuclear effects, already for densities of the order of 0.6 $\rho_0$.

\section*{Conclusion.}
      In conclusion we have studied the evolution of the quark condenaste of
pionic origin under the simultaneous influence of the baryonic density and
temperature. It is related to the scalar pionic density which comes on the one
hand from the virtual nuclear pions and on the other hand from the thermally
excited ones. We have expressed the first contribution in terms of the nuclear
response to a pion-like excitation and evaluated it for the case of nuclear
matter in the RPA scheme, first at zero and then at finite temperature. We have
shown that the RPA produces a sizeable enhancement ($\approx 30\%$), while
instead the temperature washes out the peaks and suppresses the nuclear
response, hence decreasing the virtual pion density.

  As for the thermally excited pions we have shown that the presence of the
  baryonic background appreciably enhances their number. The cause has to be
  found in the lowering of the quasi-pion excitation energies, which favours
  their thermal excitation. When the densities of both types of pions are added,
  the mutual influences which go in opposite directions cancel their effects to
  a large extent. In the density and temperature domain that we have explored, 
  the additive assumption of Eq.~(\ref{addit}) which neglects the mutual influence 
  is a good approximation. It deviates from the exact result by no more than
  15\%, the deviation being maximum around $T \approx 50$ MeV. At this T value
  the additive approximation slightly overestimates the pionic density.
  
   Our study has shown that, even at moderate baryonic density, the virtual 
   nuclear pions are a major component of the overall scalar pion density. As an
   example, at nuclear matter density, they dominate in the temperature range we
   have considered, {\it i.e.} up to at least $T \approx 150$ MeV. Since the
   pion is the agent for the mixing of the vector and axial correlators, the
   consequence of our study is that the existence of a baryonic background, if
   any, should not be ignored in this mixing.

\newpage


\end{document}